# Synthesis and optoelectronic properties of CdSe quantum dots

Hassan Ghasemi and Mohammad Hazhir Mozaffari [*]

Department of Electrical Engineering, Sanandaj Branch, Islamic Azad University, Sanandaj, Iran

[*] **Corresponding author: mh.mozaffari@iausdj.ac.ir**

**ABSTRACT**— A series of Cadmium Selenide nanoparticles in the range of 6-12 nm have been synthesized by a wet chemical route that employs Mercaptoethanol as a capping agent. These nanoparticles have been characterized by Ultraviolet-visible (UV-VIS) absorption, photoluminescence spectroscopy, and X-RAY diffraction (XRD). The effect of concentration on the optical spectra has been investigated whereas Mercaptoethanol which is a growth-limiting agent, prevents bulking. The augmentation of concentration of Mercaptoethanol decreases the size of Cadmium Selenide nanoparticles and shifts the peaks of absorption spectra.

## I. INTRODUCTION

Semiconductor nanoparticles are very different from those of the corresponding bulk materials [1]. Interesting and strongly size-dependent optical properties arise in these materials when the nanoparticle CdSe is small compared to the natural length scale of the electron-hole pair the exciton Bohr radius that calculated as:

$$a_\beta = (\hbar \mathcal{E}/e^2)(1/m_e + 1/m_h) \quad (1)$$

$$m_e = 0.13, \; m_h = 0.16, \; \mathcal{E} = 6.1$$

This radius is calculated for CdSe 54 A°, CdS 20 A°, and Zns 11 A°[2]. This range of electro-optical properties is very significant. Cdse QDS has attracted a large amount of interest owing to its exotic mechanical, thermal stability, and optical properties [3]. The unusual electronic properties lead to various applications in nanotechnology, nanoscience, nanoelectronic, and fabrication of Nanodevices. One of the methods for producing nanoparticles is the liquid state method The chemical control method is its component. The first effect of reducing size is increasing the ratio of the surface to the volume of nanoparticles CdSe-derived nanoparticles with sizes below 10 nm exhibit a property known as quantum confinement. Quantum confinement results when the electrons in a material are confined to a very small volume [4]. Quantum confinement is size-dependent, meaning the properties of CdSe nanoparticles are tunable based on their size [5]. One type of CdSe nanoparticle is a CdSe quantum dot. This discretization of energy states results in electronic transitions that vary by quantum dot size. Larger quantum dots have closer electronic states than smaller quantum dots which means that the energy required to excite an electron from HOMO to the LUMO is lower than the same electronic transition in a smaller quantum dot [6-8]. This quantum confinement effect can be observed as a redshift in absorbance spectra for nanocrystals with larger diameters. CdSe quantum dots have been implemented in a wide range of applications including solar cells, nanocrystal, quantum dot laser [9], chemical sensor [10], medical, light-emitting diodes, and fluorescent biosensing [11]. CdSe-based materials also have potential uses in biomedical imaging. Human tissue is permeable to near infra-red light [12]. By injecting appropriately prepared CdSe nanoparticles into injured tissue, it may be possible to image the tissue in those injured areas. optoelectronic and photonic devices may directly benefit from the unique characteristics of quantum dots especially their discrete density of state compared to bulk.In recent papers ,Cdse nanoparticles have been synthesized using a variety of methods that involve solid-state reaction photochemical method [13], and electrochemical method.

## II. EXPERIMENT

### A. *synthesis of m.c.e capped CdSe Nanoparticles*

Inside a three-span balloon .587 g of Cadmium acetate dihydride with 50 ml of dimethyl sulfate, with a concentration of 2.2 mM, and uniformly dispersed with a magnetic stirrer for 10 minutes. Then, add a different amount of concentration of m.c.e (take a drop every 3 seconds). In the last step, sodium selenide hydride dissolved at a concentration of 2.9 mM in 8 ml of water, and immediately after the completion of ether sulfate, Add/ drop drops and, when the drops of sodium selenide Penta hydride are completed, the whole solution is rotated with a magnetic stirrer for the first half-hour, with the same initial speed. Turn off the magnetized agitator and then turn on the heater (raise the temperature) When the temperature reached 80 ° C, the solution was refluxed for three hours and then the solution was washed with acetone and after the half, The clock at the end is also 5 to 7 time.

### B. *Instruments*

Power XRD patterns were made on an X Pertpro Philips Spectrometer using Cu kα radiation (wavelength of 1.541 A). The photoluminescence spectrum was acquired on LS-55 company Perkin Elmer. the UV-Vis spectrum was measured on Shimadzu UV-3101s pc spectrophotometer.

## III. RESULTS AND DISCUSSION

### A. *UV –Vis absorption spectrum*

Fig 1 shows absorption spectra of m.c.e stabilized CdSe nanoparticles with different sizes separated from the corresponding crude solutions for three various contenting of m.c.e. It was observed that the absorption spectra of the samples taken were a redshift of 294 to 508 nm and their absorption peak was smaller than bulking CdSe (714 nm) particles size is calculated by approximating the effective mass and Brass equation according to data coming from table1.

$$E_g^{nano} - E_g^{bulk} = (\pi^2 \hbar^2 / 2R^2)(1/m_e + 1/m_h)$$

$$E_g^{nano} = 1240/\lambda_{max} \quad (2)$$

$$E_g^{bulk} = 1.74 \text{ eV}$$

Table 1. comparison of size nanoparticles CdSe by dissolve concentration

| Sample | a | b | c |
|---|---|---|---|
| Concentration (mM) | 0.01 | 0.1 | 0.2 |
| Absorption (nM) | 508 | 401 | 294 |
| Gap energy (eV) | 2.44 | 3.09 | 4.21 |
| Particle Radius size (nM) | 12 | 8 | 6 |

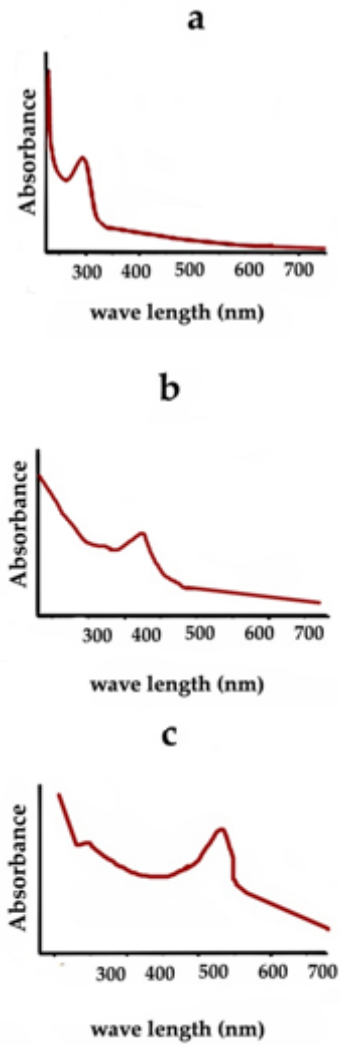

Figure1. Absorption spectra of CdSe nanoparticles (a) For concentration to Equals .01 (b) For concentration Equals to. 1 (c) For concentration Equals to .2

Table.1 indicates that an increase of gap energy leads to a decrease in the size of nanoparticles by increase the concentration of m.c.e.

## B. XRD analysis

XRD Patterns of m.c.e capped CdSe nanoparticles for sample (b) are shown in Fig.1. As seen cdse samples exhibit three peaks at 2θ= 27,46 and 51 that can be assigned to (111) and (220)and (311) plans of cubic CdSe structure .moreover the average particle size of CdSe was calculated between 5 to 12 nm using the Debye-Scherrer equation.

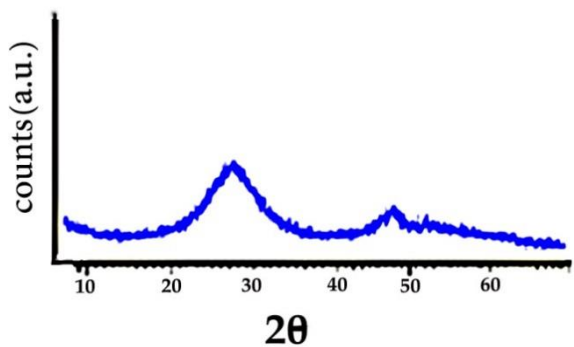

Figure. 2. XRD Pattern of the m.c.e capped CdSe nanoparticles for sample (b).

## C. PL spectrum

The material that was used for testing was diluted colloid. A photoluminescence device with a certain wavelength (close to the absorption peak) induces the material and is excitation spectra are caught by another part of the device and are sent to a computer. (A blue laser light is illuminated into a sample. The sample is started to be rolled up and completely illuminated and sent to an oscillator or detector that the product is spectrum PL). The (a) sample was captured by the Pl spectrum absorption spectrum, which is shown in Fig3.

It also shows a wavelength of absorbance of about 350 nm, with a change in the distribution wavelength in the given as default to the device. Simultaneously with the device Pl, the absorption and emission graphs of nanoparticles cdse were measured for sample (a).

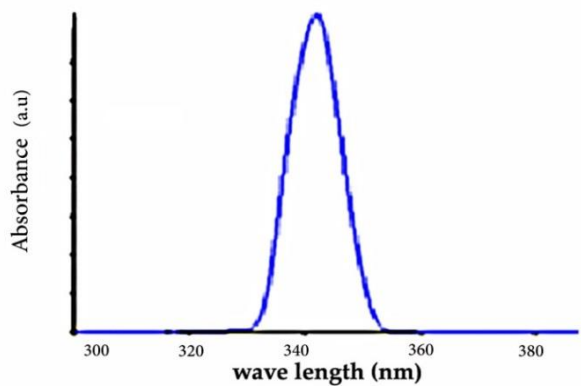

Figure. 3. Absorption spectra of CdSe nanoparticles (a) Which is measured by the pl device.

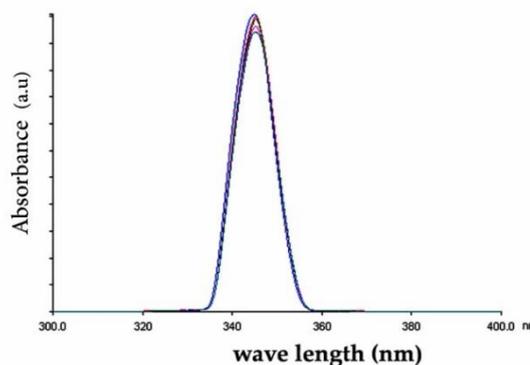

Figure. 4. Photoluminescence spectra of the m.c.e capped CdSe nanoparticles.

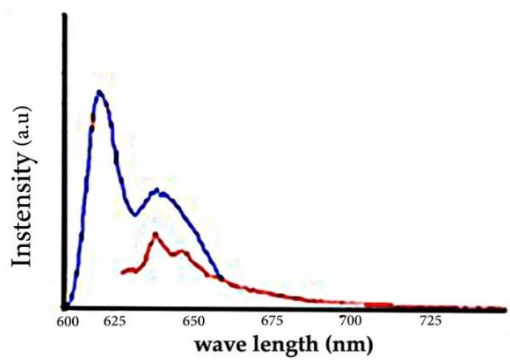

Figure. 5.Photoluminescence spectra of the m.c.e capped CdSe nanoparticles.

The maximum emission wavelength is at 625 and 640 nm. in comparison with that of bulk CdSe at 730nm. We have a blue shift in Photoluminescence spectra. the excitation wavelength was 390 This effect might be related to the quantum effects.

## IV. CONCLUSION

Particles can be produced in different sizes by wet chemicals the advantage of this method is simple, comfortable, and cheapness. CdSe nanoparticles have been produced in size 6-12 nm. All nanoparticles were in size quantization regime in comparison with bulk CdSe. The size-dependent shift of the CdSe nanoparticles bandgap energy. The spectrum of nanoparticles produced increases the bandwidth for the quantum imprisonment of electrons. The X-ray diffraction pattern shows cubic CdSe structure lattice samples for these nanoparticles. The selenide cadmium particles have a good spectrum and show that the sample has no crystal defects and has a good constructional quality.